# Large second harmonic generation of LiCs$_2$PO$_4$ caused by the metal-cation-centered groups

Xiyue Cheng[1], Myung-Hwan Whangbo[1,2,*], Guo-Cong Guo[1], Maochun Hong[1], and Shuiquan Deng[1,*]

[1] *State Key Laboratory of Structural Chemistry, Fujian Institute of Research on the Structure of Matter, Chinese Academy of Sciences, Fuzhou, Fujian 350002, China*

[2] *Department of Chemistry, North Carolina State University, Raleigh, NC 27695-8204, USA*

*Correspondence to: sdeng@fjirsm.ac.cn, mike_whangbo@ncsu.edu

**Abstract:** We evaluated the individual atom contributions to the second harmonic generation (SHG) coefficients of LiCs$_2$PO$_4$ (LCPO) by introducing the partial response functionals on the basis of first principles calculations. The SHG response of LCPO is dominated by the metal-cation-centered groups CsO$_6$ and LiO$_4$, not by the nonmetal-cation-centered groups PO$_4$ one expects from the existing models and theories. The SHG coefficients of LCPO are determined mainly by the occupied orbitals O-2p and Cs-5p as well as by the unoccupied orbitals Cs-5d and Li-2p. For the SHG response of a material, the polarizable atomic orbitals of the occupied and the unoccupied states are both important.

**Keywords:** Second Harmonic generation • LiCs$_2$PO$_4$ • Individual atom contribution • Partial response functional

Deep-ultraviolet (wave length λ < 200 nm) nonlinear optical (NLO) materials can produce coherent ultraviolet (UV) light through laser frequency conversion. Thus, they are of great interests in practical applications such as modern manufacturing, laser medical treatment and communications as well as in fundamental research [1]. Recently, there were two independent reports [2] on a new deep-UV NLO compound LCPO, made up of polar groups PO$_4$, LiO$_4$ and CsO$_6$ (**Figure 1**). It has the largest SHG response in the phosphate family (i.e., 2.6 [2a] and 1.8 [2b] times that of KH$_2$PO$_4$, KDP). Concerning the cause for this large SHG effect, the two studies have put forward conflicting explanations. Shen et al. [2b] considered the aligned PO$_4$ groups as responsible for the large SHG response based on the anionic group theory[1a]. In contrast, Li et al. [2a] considered this theory as insufficient for LCPO, suggesting that the SHG effect arises from the preferred spatial orientation of nonbonding O-2p orbitals. Qualitatively, the SHG phenomenon is discussed by considering either bond dipole moments of covalent bonds (e.g., the P-O bonds in LCPO) [1a, 3] or specific acentric atom displacements [4]. The valence electronic density distribution of a noncentrosymmetric material is determined not only by covalent bonds but also by ionic bonds (e.g., the Li-O and Cs-O bonds in LCPO). Therefore, the possibility that ionic bonds can also contribute to the SHG response cannot be excluded. Furthermore, the virtual optical excitations leading to the SHG phenomenon of a NLO material involves both the occupied and the unoccupied states of the material. (Hereafter, "virtual optical excitations" will be simply referred to as "optical excitations".) Static dipole moments are a property associated with only the occupied states in the absence of the electric field ***E*** of light. In contrast, the SHG coefficients are the second-order functional derivative of the electric polarization ***P*** with respect to ***E***. It is desirable to have a conceptual tool with which to analyze the SHG at the atomic length scale.

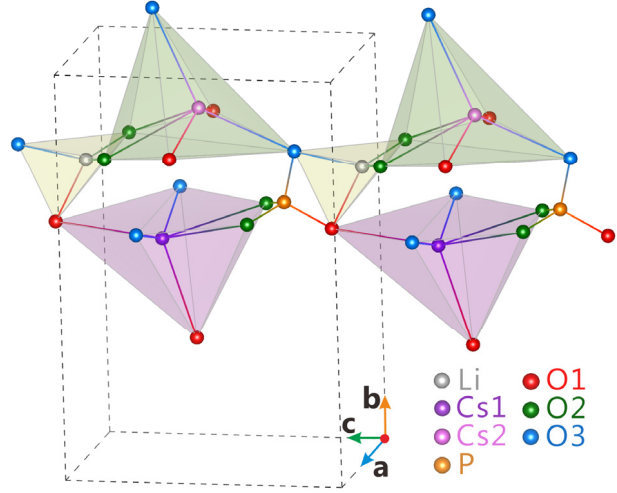

**Figure 1.** Perspective view of the "Cs1O$_6$-PO$_4$" and "Cs2O$_6$-LiO$_4$" chains running along the c-direction in a plane parallel to the bc-plane, with emphasis on the cation-centered polyhedra CsO$_6$ and LiO$_4$. A layer parallel to the bc-plane has four chains per unit cell, but only two chains are shown for clarity. These chains are condensed by sharing oxygen atoms to form a three-dimensional network (for details, see the supplementary information **S1**, **Figure S1**). The arrangements of the cations around the O$^{2-}$ anions are shown in **Figure S2**.

In the present work the SHG coefficients of LCPO are examined by using first-principles calculations. To determine the contributions of specific atomic orbital states, and hence the relative contributions of the individual atoms, to the SHG coefficients, we introduce the partial response functionals for the SHG coefficients associated with the valence bands (VBs) and the conduction bands (CBs). We find that the strong SHG response of LCPO results not only from the anions O$^2$ but also from the cations Cs$^+$ and Li$^+$, revealing that the CsO$_6$ and LiO$_4$ groups contribute much more to the SHG response of LCPO than do the PO$_4$ groups.

LCPO, crystallizing in the noncentrosymmetric space group $Cmc2_1$, has three crystallographically nonequivalent O atom sites (O1, O2 and O3), two crystallographically nonequivalent Cs atom sites (Cs1 and Cs2), as well as unique Li and P atom sites. With the Cs-O distance of ~3.14 Å [5], the Cs$^+$ ions form the Cs1O$_6$ and Cs2O$_6$ pentagonal pyramids. The arrangement of the cation-centered polyhedra (i.e., LiO$_4$, Cs1O$_6$, Cs2O$_6$ and PO$_4$) in LCPO is presented in **Figure 1**. All cation-centered polyhedra are polar, and so are the anion-centered polyhedra. The arrangements of the Li-O, Cs1-O, Cs2-O and P-O bond dipole moments in LCPO lead to a nonzero static dipole moment only along the c-direction (for details, see the supplementary information, **S1**).

The SHG phenomenon of a polar material arises from how the electrons of the material respond to the oscillating electric field $E$ of light, which can be described by the electric polarization $P$ expressed as,

$$P = P^{(0)} + P^{(1)} + P^{(2)} + \cdots, \quad (1)$$

where $P^{(0)}$, $P^{(1)}$ and $P^{(2)}$ are the zero-, first- and second-order polarizations, respectively. $P^{(0)}$ is independent of $E$, with component $P_i^{(0)}$ ($i = x, y, z$). With $\epsilon_0$ denoting the electric permittivity of the vacuum and using the Einstein summation convention, the $i^{th}$ components of $P^{(1)}$ and $P^{(2)}$ (in the domain of frequency $\omega$) can be

written as

$$P_i^{(1)}(\omega) = \epsilon_0 \chi_{ij}^{(1)}(\omega) E_j(\omega), \tag{2}$$

$$P_i^{(2)}(\omega) = \epsilon_0 \chi_{ijk}^{(2)}(\omega) E_j(\omega_1) E_k(\omega_2). \tag{3}$$

In Eq. 2, $\omega$ refers to the frequency of the pump light as well as that of the response signal. In Eq. 3, $\omega$ is the frequency of the response signal while $\omega_1$ and $\omega_2$ are those of the pump lights with $\omega = \omega_1 + \omega_2$. The linear electric susceptibility $\chi^{(1)}$ is a second-rank tensor, while the nonlinear electric susceptibility tensor (or the SHG coefficient tensor) $\chi^{(2)}$ is a third-rank tensor. $\chi^{(1)}$ and $\chi^{(2)}$ are first- and second-order functional derivatives of $P$ with respect to $E$, while the zero-order (or spontaneous) polarization $P^{(0)}$ has no dependence on the electric field. Once the space group of a material is known, what components of its $\chi^{(2)}$ are nonzero is dictated by the crystal symmetry and other symmetries governing the susceptibility tensors. To understand why a NLO material has a particular SHG response, it is necessary to analyze its symmetries to find nonzero SHG coefficients $\chi_{ijk}^{(2)}$, calculate their values, and then explore how they are related to the electronic structure of the material.

We begin our quantitative study of the SHG of LCPO based on density functional theory (DFT) by first optimizing its crystal structure using the Vienna *ab-initio* Simulation Package (VASP) [6]. This results in a crystal structure very similar to the experimental one (see **S2** and **Table S1**). The optimized structure was used in all our calculations at various levels of the VASP and ABINIT [7] (see **S3** and **S4**). In our study, the Cartesian x-, y- and z-axes are taken along the crystallographic a-, b- and c-directions of LCPO, respectively. The bandgaps obtained by DFT calculations were corrected with the scissor operation [8] or by the hybrid functional HSE06 calculations [9] with mixing parameter $\alpha = 0.4$. We also carried out the GW calculations (see **S3**). The optical properties of LCPO were calculated by using the "sum over states (SOS)" and the density functional perturbation theory (DFPT) methods in the ABINIT code after employing the scissor operation (see **S5** and **S6**). As a tool for analyzing the contributions of the atomic orbital states to the SHG coefficients, we introduce the partial response functionals for the VBs and CBs and evaluate them as a function of either the energy $E$ or the band index $I$ (see **S7**).

In calculating the susceptibilities $\chi^{(1)}$ and $\chi^{(2)}$ of a material, use of a correct bandgap is necessary. With the GGA [10] calculations, the band gap of LCPO is calculated as $E_g^{PBE} = 4.43$ eV, much smaller than the experimental value $E_g^{exp} = 7.02$ eV [2a]. In computing optical properties, this deficiency of the DFT [11] is often corrected empirically by employing the scissor operation [8], in which the CBs are shifted in energy to have the experimental bandgap [12]. The GW calculations for LCPO at the GW$_0$ level [13] gives a bandgap $E_g^{GW} = 7.06$ eV, in good agreement with experiment. Since the formalism for calculating the optical properties is based on the single particle approximation [14], we employ the HSE06 [9, 15] calculations to determine the electronic and optical properties. With the default mixing parameter $\alpha = 0.2$, the HSE06 calculations give the bandgap $E_g^{HSE} = 6.05$ eV. The HSE06 calculations with $\alpha = 0.4$ gives the bandgap $E_g^{GW} = 7.06$ eV (the inset of **Figure S4**), and is used in calculating all the electronic properties reported below.

To know what occupied and unoccupied states of LCPO are involved in the excitations leading to its SHG response,

we first examine the electronic band structure calculated for LCPO (**Figure S4**). The VBs are considerably narrower than the CBs, reflecting that the O-2s/2p, the P-3s/3p and the Cs 5p orbitals forming the VBs are contracted, whereas the Li-2s/2p and the Cs-5d/6s orbitals forming the CBs are diffuse. Our COHP [16] calculations show that the covalent character is strong in the P-O bonds, weak in the Li-O bonds, and nearly absent in the Cs-O bonds (**Figure S5**). The analysis of the projected density of states (PDOS), shows that the energy range between -10.0 and 25.0 eV can be divided into five regions I – V (see below and **Figures S6**-**S7** for more details).

The results of our calculations on the $P^{(0)}$ and $\chi^{(1)}$ (see **S5** and **Table S2-S4**) show that the first-order property $\chi^{(1)}$ is practically isotropic (**Figures S8-S10**), in contrast to the highly anisotropic zero-order property $P^{(0)}$. The second-order electronic susceptibility $\chi^{(2)}$ has 27 components, $\chi^{(2)}_{ijk}$. The application of symmetries leads to only three nonequivalent components (see **S6**). We use the Voigt matrix notation [17] $d_{ij}$ to represent the symmetry-reduced third-rank tensor $\chi^{(2)}_{ijk}$ (see **S6.1**). The nonzero $\chi^{(2)}_{ijk}$ components occur in all three Cartesian directions, so the second-order property is nearly isotropic like the first-order property. The three independent nonzero components $d_{31}$, $d_{32}$ and $d_{33}$ were calculated by using six different methods, Methods 1 – 6, are summarized in **Table 1**.

**Table 1**. SHG components $d_{ij}$ (ij = 31, 32, 33) and $d_{eff}$ obtained by six different computational methods

|   | Method | $d_{31}$ | $d_{32}$ | $d_{33}$ | $d_{eff}$ |
|---|---|---|---|---|---|
| 1 | GGA-PAW | -1.281 | 0.530 | 1.130 | 0.87 |
| 2 | LDA-TM (DFPT) | -1.129 | 0.326 | 1.223 | 0.79 |
| 3 | HSE06-PAW | -1.056 | 0.443 | 0.937 | 0.72 |
| 4 | GGA-ONCV ($\omega$ = 1.17 eV) | -0.979 | 0.400 | 0.969 | 0.68 |
| 5 | GGA-ONCV | -0.695 | 0.281 | 0.684 | 0.48 |
| 6 | LDA-TM (DFPT) $E_g$ = 4.43 eV | -1.83 | 0.64 | 2.05 | 1.32 |

[a] VASP calculations for Methods 1 [10, 18] and 3 [9a]; ABINIT calculations for Methods 2 [19], 4 [20], 5 [20] and 6 [19].
[b] Use of $E_g$ = 7.06 eV for Methods 1 – 5.
[c] Non-static calculations ($\omega$ = 1.17 eV) for Method 4; Static limit calculations ($\omega$ = 0) for Methods 1 – 3, 5 and 6.
[d] SOS calculations for Methods 1, and 3 – 5; DFPT calculations for Methods 2 and 6.
[e] GGA = Generalized Gradient Approximation, LDA = Local-Density Approximation, ONCV = Optimized Norm-Conserving Vanderbilt, PAW = Projector Augmented-Wave, TM = Troullier-Martins.

The corrected bandgap (i.e., 7.06 eV) was used in Methods 1, 2, 4 and 5 by using the scissor operation, and in Method 3 by using HSE06 ($\alpha$ = 0.4). Method 6 is the same as Method 2 except that the scissor operation was not included, so a smaller bandgap (i.e., 4.43 eV) was used. To compare with the experimental values, we calculated the effective, $\chi_{eff}$, from the $\chi^{(2)}_{ijk}$ values by using the Kurtz-Perry method [21] and finally presented them in terms of the Voigt notation $d_{eff} = 1/2\, \chi_{eff}$. The experimental $d_{eff}$ values reported for LCPO are somewhat different [i.e., 0.86 pm/V =

~2.6×$d_{\text{eff}}^{\text{KDP}}$ [2a] and 0.60 pm/V = ~1.8×$d_{\text{eff}}^{\text{KDP}}$ [2b], see **S6.3**]. When the corrected bandgap is used (Methods 1 – 5), the calculated $d_{\text{eff}}$ values range from 0.48 to 0.87 pm/V (**Table 1**), which are comparable in magnitude and in good agreement with the experimental values. In calculating the $d_{ij}$ values, use of a smaller bandgap (Method 6) causes the largest error and the differences in other factors affecting electronic structures (Methods 1 – 5) do not exert a strong influence (see **S6.2** and **S7**).

In general, second-order NLO properties are determined by the optical excitations from the VBs to the CBs via the intermediate states covering both the VBs and CBs (see **S6.2**). Suppose that, as depicted in **Figure 2a**, the VBs range from $E_{\text{min}}$ to the valence band maximum, VBM, while the CBs range from the conduction band minimum, CBM, to $E_{\text{max}}$.

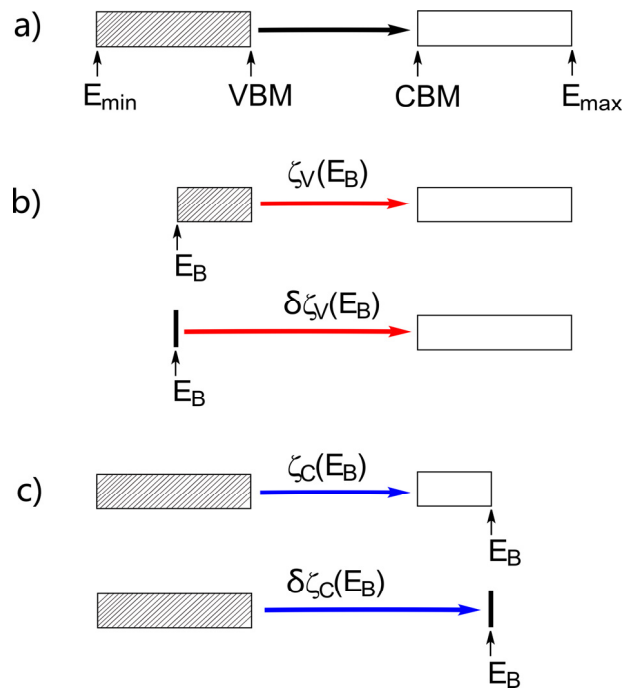

**Figure 2.** Concepts of the partial SHG functionals: a) the total response, b) the partial responses from the VBs, $\zeta_V(E_B)$ and $\delta\zeta_V(E_B)$, and c) the partial responses from the CBs, $\zeta_C(E_B)$ and $\delta\zeta_C(E_B)$.

The total response to the second-order NLO property is determined by the excitations from all occupied states of the VBs to all unoccupied states of the CBs via the intermediates states. The contribution of a certain occupied energy region between $E_B$ and VBM, $\zeta_V(E_B)$, to the overall optical property is determined by the excitations from all occupied states between $E_B$ and VBM to all the unoccupied states of the CBs (**Figure 2b**), and the contribution, $\delta\zeta_V(E_B)$, of a specific occupied state of energy $E_B$ to the overall optical property by the excitations from that occupied state to all unoccupied states of the CBs (**Figure 2b**). Similarly, the contribution, $\zeta_C(E_B)$, of a certain unoccupied region between CBM and $E_B$ to the overall optical property is determined by the excitations from all occupied states of the VBs to all unoccupied states between CBM and $E_B$ (**Figure 2c**), and the contribution, $\delta\zeta_C(E_B)$, of a specific unoccupied state of energy $E_B$ to the overall optical property by the excitations from all occupied states of the VBs to that unoccupied state (**Figure 2c**). The quantitative evaluations of these quantities $\zeta_V(E_B)$, $\delta\zeta_V(E_B)$, $\zeta_C(E_B)$, and

$\delta\zeta_C(E_B)$, are discussed in detail in **S7.1.**

The partial response functionals evaluated as a function of $E_B$ for $\chi^{(2)}_{311}$, $\chi^{(2)}_{322}$ and $\chi^{(2)}_{333}$ are summarized in **Figure 3**. We first analyze how these coefficients vary by examining the $\zeta_V(E_B)$ functionals (**Figure 3a**). In region I where the nonbonding O-2p states dominate (**Figure 3e**), these coefficients increase in magnitude with decreasing $E_B$ from the VBM. In region II where the P-O bonding states occur (**Figure 3e**), the three coefficients do not change much, indicating that the contribution of the PO4 groups to the SHG is small. In region III where the Cs-5p orbital states dominate, the $\chi^{(2)}_{333}$ increases substantially, but the $\chi^{(2)}_{311}$ and $\chi^{(2)}_{322}$ change little. Let us now examine how the coefficients $\chi^{(2)}_{311}$, $\chi^{(2)}_{322}$ and $\chi^{(2)}_{333}$ vary with increasing $E_B$ above the CBM by examining the $\zeta_C(E_B)$ functions (**Figure 3b**). These coefficients increase in magnitude steadily in region IV where the Cs-5d states dominate and make antibonding with O-2p states. In region V where the Li-2p states contribute strongly but the Cs 5d states still dominate, the three coefficients increase in magnitude more strongly compared with those of region IV. In short, the occupied states important for the SHG of LCPO are the O-2p states of region I, and the Cs-5p states of region III, while the unoccupied states important for the SHG are the Cs-5d states of region IV and V as well as the Li-2p states of region V. The importance of the Cs-5p and Cs-5d states as well as the Li-2p states can be readily identified by comparing the PDOS plots (**Figure 3e**) with the $\delta\zeta_V(E_B)$ and $\delta\zeta_C(E_B)$ functionals (**Figure 3c,d**).

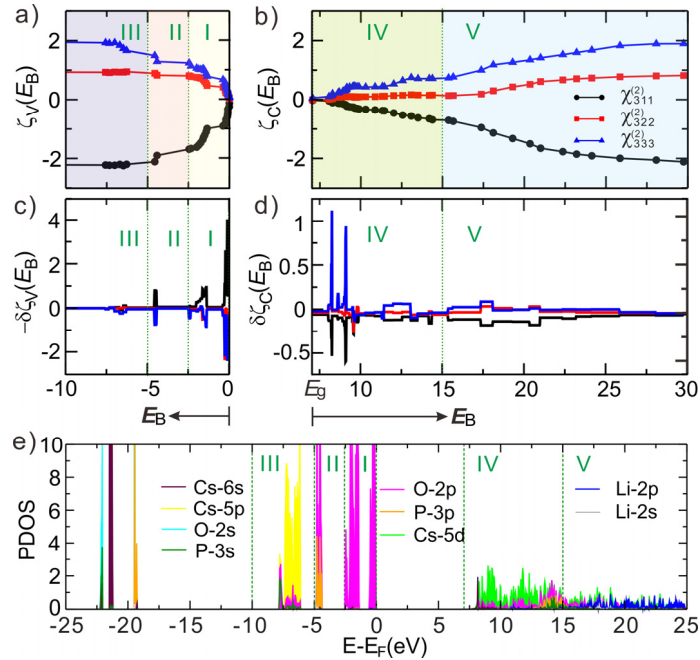

**Figure 3.** a) $\zeta_V(E_B)$-vs-$E_B$ plot, b) $\zeta_C(E_B)$-vs-$E_B$ plot, c) $-\delta\zeta_V(E_B)$-vs-$E_B$ plot, and (d) $\delta\zeta_C(E_B)$-vs-$E_B$ plot. The values of the functionals are in pm/V, $2d_{31} = \chi^{(2)}_{311}$, $2d_{32} = \chi^{(2)}_{322}$, and $2d_{33} = \chi^{(2)}_{333}$.

The above observations reflect that optical excitations are largely local in nature because the transition dipole moment is a local operator (see **S6.2** and **S7**). Each $O^{2-}$ anion is coordinated with the surrounding cations (**Figure S2**), so its O-2p lone pair states have some contributions of the cation orbitals (e.g., Cs-5d, Li-2p, P-3p). Similarly, each cation ($Cs^+$, $Li^+$, $P^{5+}$) is coordinated with the surrounding $O^{2-}$ anions (**Figure 1**), so its unoccupied states have some contributions of the anion O-2p orbitals. The perturbation of the oscillating light field $E$ induces a dynamic mixing between the occupied and unoccupied states involving the $O^{2-}$ anions. This dynamically affects the electron distributions surrounding both the O and their surrounding cations. That is, the SHG response involves a dynamic polarization of the electron cloud.

To evaluate the individual atom contributions to the SHG components, $d_{ij}$, it is convenient to express the corresponding partial response functionals in terms of the band index $I_B$, $\zeta(I_B)$, where the band index $I_B$ runs from 1 to $N_{tot}$ with increasing the energy, $E_B$, from $E_{min}$ to $E_{max}$. Then, the functionals $\zeta_V(I_B)$ and $\zeta_C(I_B)$ as well as the derivative functionals $\delta\zeta_V(I_B)$ and $\delta\zeta_C(I_B)$ can be obtained as described in **S7.1** and **Figure S11-S12**.

The contribution, $^{VB}A_\tau$ ($^{CB}A_\tau$), each individual atom $\tau$ makes to the SHG response from the VBs (CBs) is summarized in **Tables S5 (S6)**. The total contribution $A_\tau$, each individual atom $\tau$ makes to the SHG response is given by $A_\tau = (^{VB}A_\tau + {}^{CB}A_\tau)/2$ (**Table 2**). The relative atom contributions decrease in the order O >> Cs > P > Li in the VB contributions (**Table S5**), and in the order Cs > Li > P > O in the CB contributions (**Table S6**). These findings reflect that the SHG of LCPO is governed largely by the occupied orbital states O-2p and Cs-5p and by the unoccupied orbital states Cs-5d and Li-2p. In terms of both the VB and CB contributions, the relative atom contributions decrease in the order O > Cs >> Li > P (**Table 2**). The metal-cations of $LiO_4$ and $CsO_6$ are much more important than the nonmetal cation of $PO_4$ in determining the strength of the SHG response.

**Table 2.** The contributions of the Li, Cs, P and O atoms to the SHG components $d_{31}$, $d_{32}$ and $d_{33}$.

| Atom | $d_{31}$ (%) | $d_{32}$ (%) | $d_{33}$ (%) |
|---|---|---|---|
| Li | 6.27 | 7.00 | 5.59 |
| Cs1 | 10.54 | 10.35 | 14.43 |
| Cs2 | 10.77 | 10.41 | 14.56 |
| P | 6.00 | 5.07 | 5.96 |
| O1[a] | 33.46 | 33.37 | 29.51 |
| O2 | 15.80 | 17.06 | 15.55 |
| O3 | 17.15 | 16.75 | 14.41 |

[a] There are two O1 atoms per formula unit.

In summary, the SHG coefficients $d_{eff}$ of LCPO, obtained from first-principles calculations, are in good agreement with experiment at various levels of calculations. The use of the partial response functionals enabled us to evaluate the individual atom contributions to the SHG response. The metal-cation-centered groups $CsO_6$ and $LiO_4$ contribute more strongly to the SHG response of LCPO than do the nonmetal-cation-centered groups $PO_4$, suggesting a strong SHG response for a NLO system with occupied and unoccupied states consisting of easily polarizable atomic orbital states.

**Acknowledgements**

This work is financially supported by the National Natural Science Foundation of China (21703251); the Strategic Priority Research Program of the Chinese Academy of Sciences (XDB20000000); National Key Research and Development Program of China (2016YFB0701001); 973 Program of China (2014CB932101); 100 talents program of CAS and Fujian Province.

# SUPPORTING INFORMATION

# Large second harmonic generation of LiCs$_2$PO$_4$ caused by the metal-cation-centered groups

Xiyue Cheng[1], Myung-Hwan Whangbo[1,2,*], Guo-Cong Guo[1], Maochun Hong[1], and Shuiquan Deng[1,*]

[1] *State Key Laboratory of Structural Chemistry, Fujian Institute of Research on the Structure of Matter, Chinese Academy of Sciences, Fuzhou, Fujian 350002, China*

[2] *Department of Chemistry, North Carolina State University, Raleigh, NC 27695-8204, USA*

*Correspondence to: sdeng@fjirsm.ac.cn, mike_whangbo@ncsu.edu

## 1. Crystal structure of LiCs$_2$PO$_4$ (LCPO) and dipole moments

To graphically show the essential crystal symmetry of LCPO and why it has a nonzero dipole moment only along the c-direction, we show in **Figure S1** three projection views of LCPO using the ball-and-stick models of the cation-centered polyhedra, LiO$_4$, PO$_4$, Cs1O$_6$ and Cs2O$_6$, which are all polar. **Figure S1a** shows a projection view along the *a*-direction of a "Cs1O$_6$-PO$_4$-Cs2O$_6$-LiO$_4$" layer parallel to the bc-plane (hereafter, ||bc Cs1O$_6$-PO$_4$-Cs2O$_6$-LiO$_4$ layer). There is no crystal symmetry that force the bond dipole arrangements of this layer to cancel out. A projection view of this layer along the *c*-direction (**Figure S1b**) shows that the ||bc Cs1O$_6$-PO$_4$-Cs2O$_6$-LiO$_4$ layer has a mirror plane of symmetry parallel to the bc-plane, so that the bond dipole moments cancel out along the *a*-direction. The projection view along the *a*-direction in **Figure S1c** shows how the ||bc Cs1O$_6$-PO$_4$-Cs2O$_6$-LiO$_4$ layers are condensed along the *a*-direction. Every two adjacent ||bc Cs1O$_6$-PO$_4$-Cs2O$_6$-LiO$_4$ layers (differing in the a-axis height by a/2) are related to each other by a 2$_1$ symmetry along the c-direction, so that the dipole moments cancel out along the *b*-direction. Alternatively, the crystal structure of LCPO can be described by using the anion-based polyhedra, as depicted in **Figure S2**, by using the ball-and-stick model. All the anion-centered polyhedra are polar.

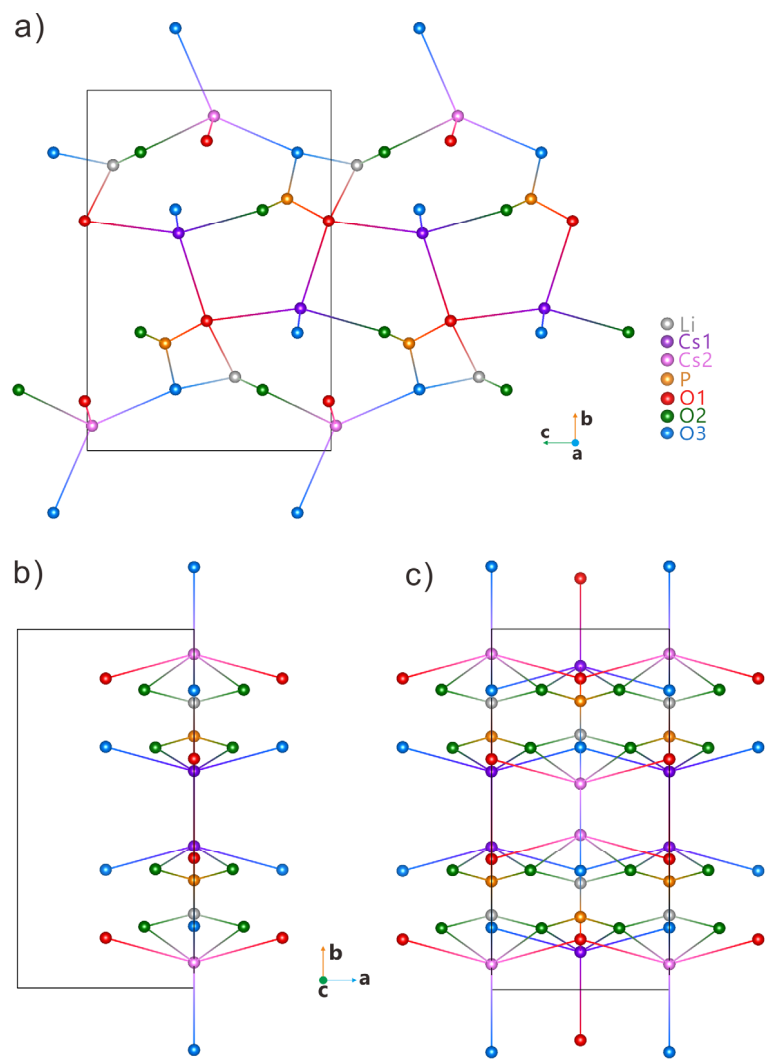

**Figure S1.** Projection views of the ||bc Cs1O$_6$-PO$_4$-Cs2O$_6$-LiO$_4$ layer along the (a) *a*-direction and (b) *c*-direction. (c) Projection view of the ||bc Cs1O$_6$-PO$_4$-Cs2O$_6$-LiO$_4$ layers are condensed to form the crystal lattice of LCPO.

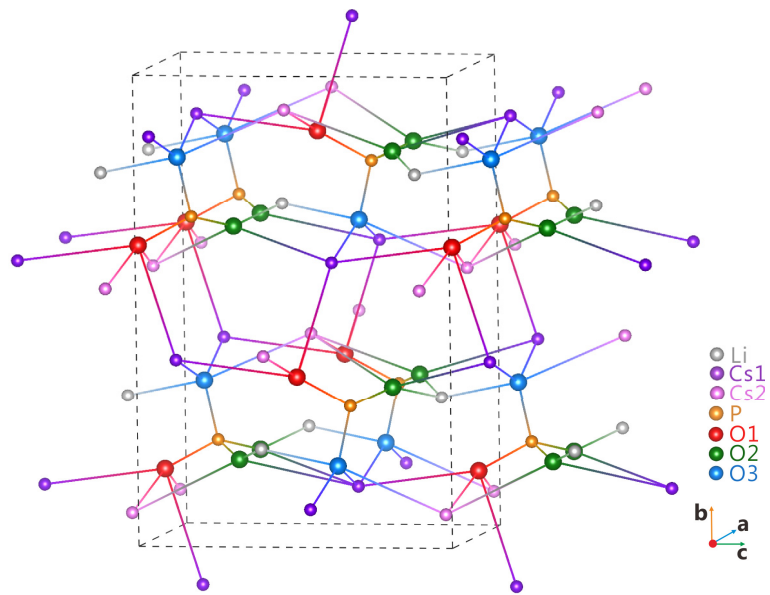

**Figure S2.** Coordination environments of the O1, O2 and O3 atoms, showing an alternative view of the crystal structure of LCPO. It emphasizes how the $O^{2-}$ anions of the O1, O2 and O3 sites are coordinated with their surrounding cations $Li^+$, $Cs1^+$, $Cs2^+$ and $P^{5+}$.

## 2. Optimized crystal structure of LCPO and dynamic stability

We optimized the orthorhombic LCPO with space group $Cmc2_1$ using the GGA/PBE calculations (see supplementary text **S3.1** for computational details). The optimized cell parameters and the optimized atom positions are presented in **Table S1**. The optimized lattice parameters are slightly larger compared with the experimental values determined at room temperature [1]. To check the dynamical stability of LCPO, we calculated the phonon dispersion curves (**Figure S3**) using the frozen phonon method by constructing the 2×2×2 supercell as implemented in the Phonopy code [2]. The phonon dispersion curves show no imaginary frequency indicating the dynamical stability of LCPO.

**Table S1.** Optimized crystal structure data for LCPO.

|  |  | Lattice parameters (Å) | | |
|---|---|---|---|---|
|  |  | *a* | *b* | *c* |
| This work |  | 5.862 | 12.382 | 8.139 |
| Exp.[1a] |  | 5.813 | 12.020 | 8.035 |
| Exp.[1b] |  | 5.811 | 11.938 | 8.007 |
| Atoms | Wyckoff sites | *x* | *y* | *z* |
| Li | 4*a* | 0.5 | 0.79 | 0.9768 |
| Cs1 | 4*a* | 0 | 0.5766 | 0.8874 |
| Cs2 | 4*a* | 0 | 0.8956 | 0.7518 |
| P | 4*a* | 0 | 0.7962 | 0.1919 |
| O1 | 8*b* | 0.7812 | 0.8256 | 0.0921 |
| O2 | 4*a* | 0 | 0.6725 | 0.2357 |
| O3 | 4*a* | 0 | 0.8573 | 0.3605 |

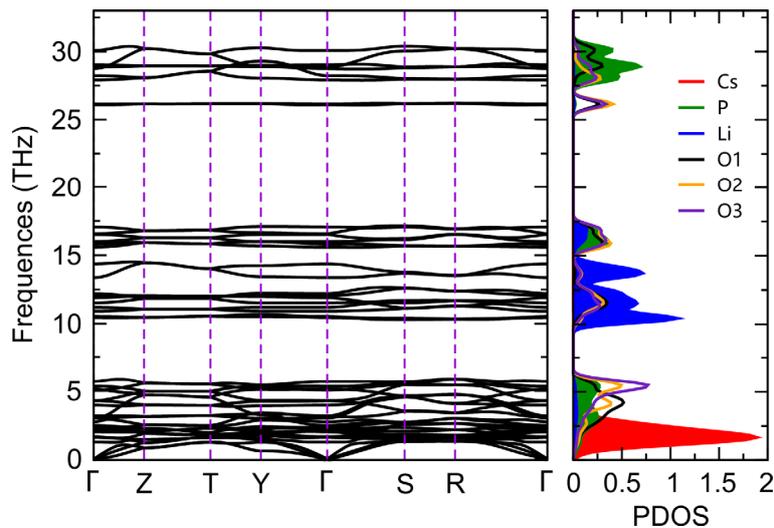

**Figure S3.** Calculated phonon dispersion curves and phonon partial DOS of LCPO.

## 3. Computational details

### 3.1. VASP calculations

**GGA-PBE calculations.** The structural and electronic properties of LCPO were calculated within the framework of density functional theory (DFT) [3] by using the Vienna ab-initio simulation package (VASP) [4] with the projector augmented wave (PAW) method [5]. The generalized gradient approximation (GGA) within the Perdew-Burke-Ernzerhof (PBE)-type exchange-correlation potentials [6] was used throughout this work. The employed PAW-PBE pseudopotentials [7] of Li, Cs, P and O treat $1s2s2p$, $5s5p6s$, $2s2p$ and $2s2p$ as the valence states, respectively. The plane wave cutoff energy for the expansion of wave functions was set at 700 eV and the tetrahedron method with Blöchl corrections was used for integrations. The numerical integrations in the Brillouin zones were performed by utilizing $7 \times 7 \times 4$ Monkhorst-Pack k-point mesh, which showed an excellent convergence of the energy differences (0.1 meV) and stress tensors (0.001 eV/ Å). The quasi-Newton algorithm as implemented in the VASP code was used in all structural relaxations. In this work, both the cell volume and the atomic positions were all allowed to relax to minimize the internal forces.

**HSE calculations.** In the HSE (Heyd-Scuseria-Ernzerhof) calculations with hybrid functional, HSE06 [8], the spatial decay of the Hartree-Fock (HF) exchange interaction is accelerated by the substitution of the bare $1/r$ Coulomb potential with a screened one [9]. The exchange-correlation energy is calculated through the hybrid functional, which mixes the HF exchange with a semi-local PBE xc function:

$$E_{xc}^{HSE} = \alpha E_x^{HF,sr}(\delta) + (1-\alpha)E_x^{PBE,sr}(\delta) + E_x^{PBE,lr}(\delta) + E_c^{PBE} \quad (1)$$

where $\delta = 0.20$ Å$^{-1}$ is a parameter controlling the separation range between the short-range (*sr*) and long-range (*lr*) parts of the Coulomb kernel, while the adjustable parameter α controls the fraction of the exact HF exchange to be incorporated.

### 3.2. GW calculations

The electron self-energy is well approximated by *GW* calculations by including many-body effects in the electron-electron interactions. We have performed the partially self-consistent $GW_0$ calculations where only the Green's function (*G*) is updated in the iteration for calculating the one-electron energy, whereas the screened Coulomb interaction ($W_0$) is fixed at the DFT level. Compared with the single-shot $G_0W_0$ calculations in which both *G* and *W* are calculated by using one electron energies and wave functions from DFT, the $GW_0$ calculations yield a better description for bandgaps [10].

### 3.3. ABINIT calculations

Calculations using the ABINIT package [11] employed the Optimized Norm-Conserving Vanderbilt (ONCV) pseudopotentials [12]. The plane-wave cutoff energy of 55 Ha and a 6×6×4 Monkhorst-Pack k-point set are found to be sufficient to reach the convergence for optical calculations. In calculating the SHG tensors using the ABINIT code, we employed the method of density functional perturbation theory (DFPT) [13] as well as the "sum over states (SOS)" method. The norm-conserving pseudopotentials generated according to the Troullier-Martins scheme [14] within the local density approximation (LDA-TM) were used in our DFPT calculations. For a comparison with the results from HSE06 calculations, we used the scissor operation to raise the bandgap to 7.06 eV (obtained by the $GW_0$ method). The scissor shift of 2.09 eV (0.0767 Ha) and 2.96 eV (0.0989 Ha) were applied to the GGA-ONCV and LDA-TM calculations, respectively.

## 4. Calculated electronic properties of LCPO

### 4.1. Band dispersion using HSE06 (α = 0.4) calculations

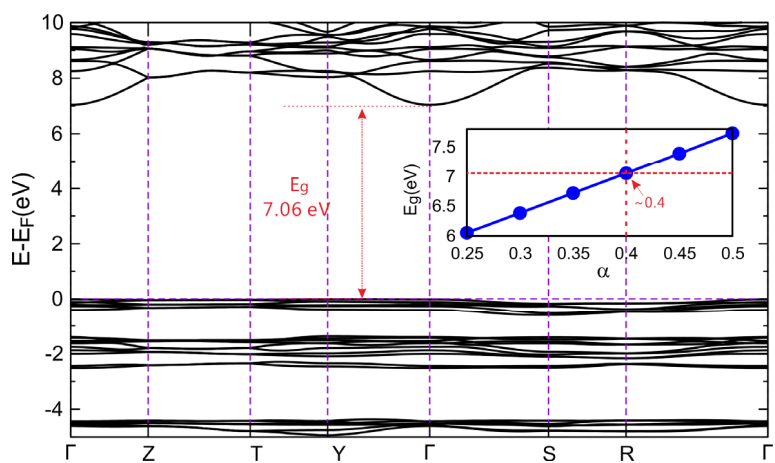

**Figure S4**. Electronic band structure of LCPO obtained by using HSE06 (α = 0.4) calculations. The inset shows how the band gap increases with increasing α.

### 4.2. Crystal orbital Hamilton population (COHP) analysis [15]

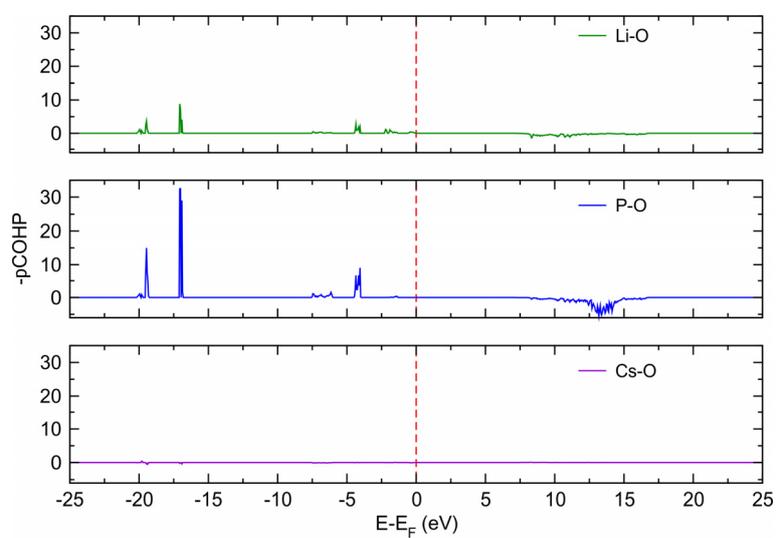

**Figure S5.** Partial COHP plots describing the Li-O, P-O and Cs-O bonding obtained from the GGA-PBE calculations with the scissor shift of 2.63 eV applied to the conduction bands.

## 4.3. PDOS using HSE06 (α = 0.4) calculations

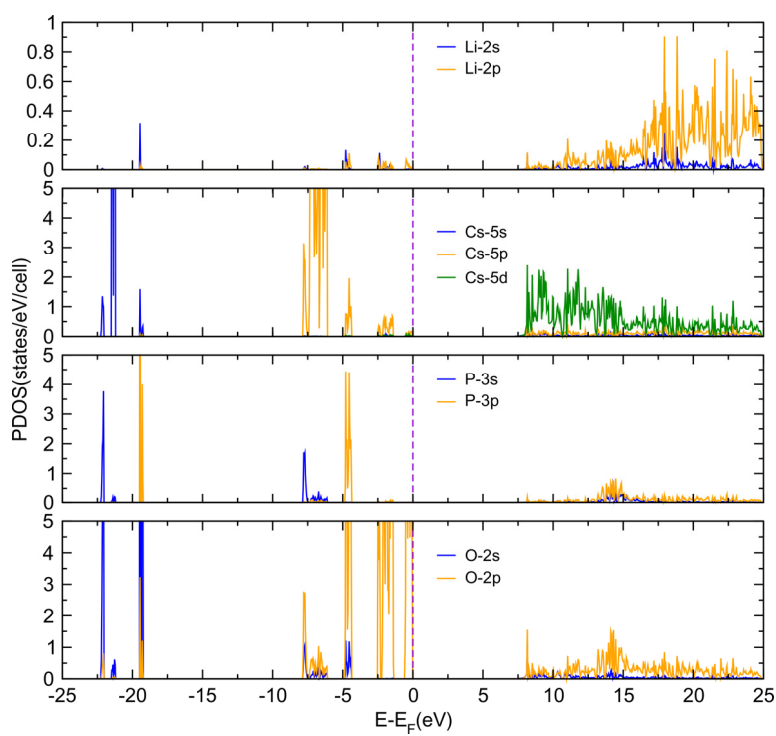

**Figure S6.** PDOS plots of LCPO obtained by using HSE06 (α = 0.4) calculations.

## 4.4. Bader charges

The Bader charge analysis [16] for LCPO was calculated using the grid-based algorithm with dense $100 \times 100 \times 100$ grids using the GGA-PBE calculations. This analysis shows that the Li, Cs and P atoms lose the valence charge by the amounts of 0.89, 0.86 and 3.60 forming $Li^{0.89+}$, $Cs^{0.86+}$ and $P^{3.6+}$ cations, respectively, whereas the O atoms gain the valence charge by the amount of 1.55 in average to create $O^{1.55-}$ anions. These results are consistent with the point charges ($Li^{1.02+}$, $Cs^{(1.04-1.12)+}$, $P^{4.78+}$ and $O^{(1.82-2.08)-}$) obtained by Li *et. al.* [1a] using the empirical bond valence calculations.

## 4.5. Partial charge density plots

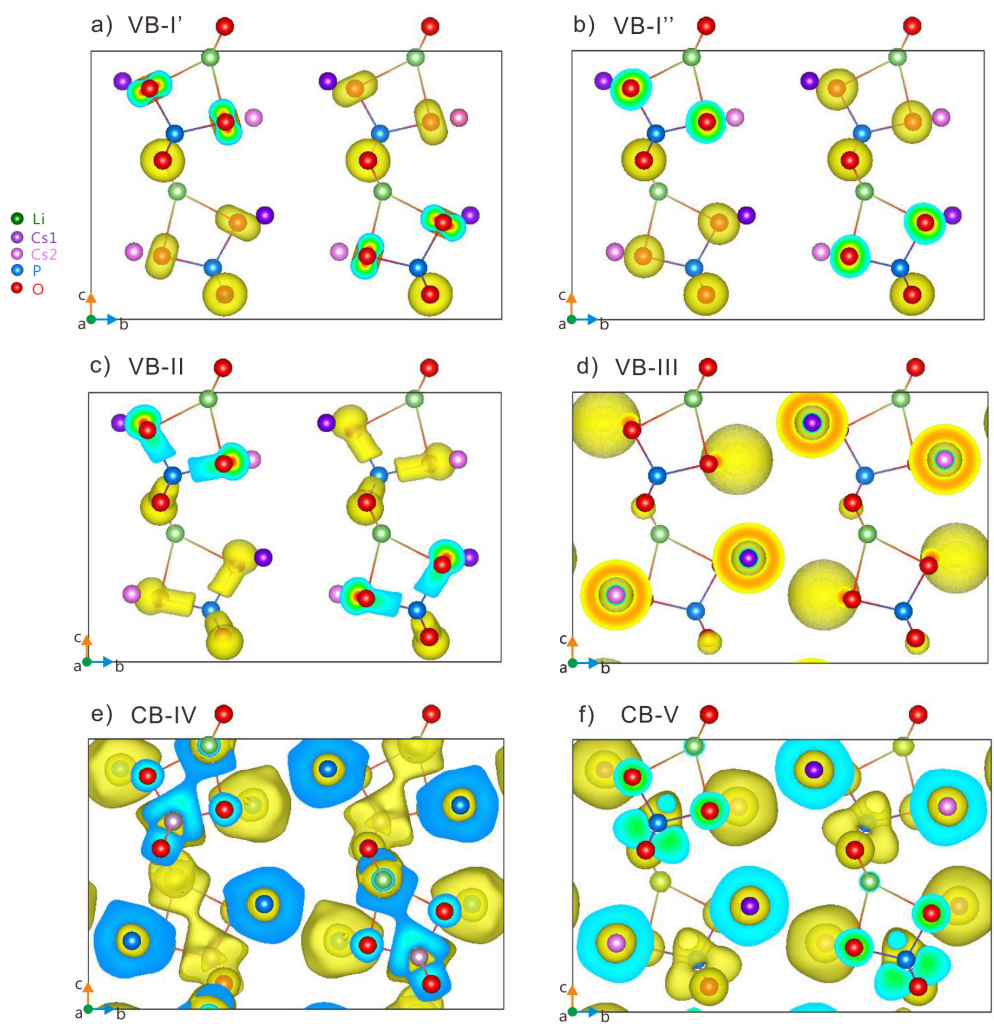

**Figure S7**. Partial charge density of LCPO obtained from the GGA-PBE calculations for the different energy regions defined in **Figure 2b** of the text. Note that the VB-I part is divided to VB-I′ (-0.85 eV to $E_F$) and VB-I″ (-2.5 to -0.85 eV), and that they represent the O-2p lone pair states.

## 5. Zero- and first-order polarizations of LCPO

### 5.1. Zero-order polarization of LCPO

We calculated the dipole moments of the $LiO_4$, $PO_4$ and $Cs1O_6$ and $Cs2O_6$ polyhedra in LCPO by using the point charges on the Li, Cs, P and O atoms derived from the Bader charge analysis (See **S4.4**), and calculate the total contributions of these polyhedra per unit cell, as summarized in **Table S2**. For comparison, the results of Li *et al.* [1a] and those of Shen *et al.* [1b] based on the empirical bond valence charges [17] are presented in **Table S3**.

**Table S2.** Dipole moments (in Debye) of $PO_4$, $LiO_4$, and $CsO_6$ polyhedra and their contributions per unit cell.

|  | *x* | *y* | *z* | Magnitude | Ref. |
|---|---|---|---|---|---|
| $LiO_4$-1 | 0.00 | -2.49 | -5.21 | 5.78 | This work |
| $LiO_4$-2 | 0.00 | 2.49 | -5.21 | 5.78 |  |
| $LiO_4$-3 | 0.00 | -2.49 | -5.21 | 5.78 |  |
| $LiO_4$-4 | 0.00 | 2.49 | -5.21 | 5.78 |  |
| $\sum LiO_4$ | 0.00 | 0.00 | -20.85 | 20.85 |  |
| $PO_4$-1 | 0.00 | 0.41 | 1.09 | 1.16 | This work |
| $PO_4$-2 | 0.00 | -0.41 | 1.09 | 1.16 |  |
| $PO_4$-3 | 0.00 | 0.41 | 1.09 | 1.16 |  |
| $PO_4$-4 | 0.00 | -0.41 | 1.09 | 1.16 |  |
| $\sum PO_4$ | 0.00 | 0.00 | 4.35 | 4.35 |  |
| $Cs1O_6$-1 | 0.00 | 15.20 | -25.60 | 29.77 | This work |
| $Cs1O_6$-2 | 0.00 | -15.20 | -25.60 | 29.77 |  |
| $Cs1O_6$-3 | 0.00 | 15.20 | -25.60 | 29.77 |  |
| $Cs1O_6$-4 | 0.00 | -15.20 | -25.60 | 29.77 |  |
| $\sum Cs1O_6$ | 0.00 | 0.00 | -102.40 | 102.40 |  |
| $Cs2O_6$-5 | 0.00 | -5.84 | 20.92 | 21.72 | This work |
| $Cs2O_6$-6 | 0.00 | 5.84 | 20.92 | 21.72 |  |
| $Cs2O_6$-7 | 0.00 | -5.84 | 20.92 | 21.72 |  |
| $Cs2O_6$-8 | 0.00 | 5.84 | 20.92 | 21.72 |  |
| $\sum Cs2O_6$ | 0.00 | 0.00 | 83.68 | 83.68 |  |
| Total | 0.00 | 0.00 | -35.22 | 35.22 | This work |

**Table S3.** Dipole moments (in Debye) of PO$_4$ polyhedra and their contributions per unit cell.

|          | x | y     | z     | Magnitude | Ref. |
|----------|---|-------|-------|-----------|------|
| PO$_4$-1 | 0 | -0.44 | 0.95  | 1.05      | Li et al. [1a] |
| PO$_4$-2 | 0 | -0.44 | 0.95  | 1.05      |      |
| PO$_4$-3 | 0 | -0.44 | 0.95  | 1.05      |      |
| PO$_4$-4 | 0 | 0.44  | 0.95  | 1.04      |      |
| PO$_4$-5 | 0 | 0.44  | 0.95  | 1.05      |      |
| PO$_4$-6 | 0 | 0.44  | 0.95  | 1.04      |      |
| $\sum$PO$_4$ | 0 | 0 | 5.7 | 6.28 |      |
| PO$_4$-1 | 0 | 0.41  | -0.99 | 1.07      | Shen et al. [1b] |
| PO$_4$-2 | 0 | 0.41  | -0.99 | 1.07      |      |
| PO$_4$-3 | 0 | -0.41 | -0.99 | 1.07      |      |
| PO$_4$-4 | 0 | -0.41 | -0.99 | 1.07      |      |

## 5.2. First-order polarization of LCPO

The two group generators, i.e., the 2-fold rotation and any one of the two mirror operations of the *mm*2 point group, enforce the second-rank tensor $\chi^{(1)}$ to be diagonal. The calculated dielectric function $\varepsilon(\omega)$ shows only three nonzero diagonal components in agreement with the symmetry conditions. Our calculations obtain a value of 0.96 for $\varepsilon_1 = Re(\varepsilon)$ at $\omega$ =109.7 eV, which is close to the asymptotic value of RPA (random phase approximation), $\varepsilon_1^{RPA}(\infty) = 1$[18]. Experimental measurements are needed to check the accuracy of the calculated value for $\varepsilon_1(\infty)$, because it is smaller than those of the typical ionic compounds (~2 - ~3)[19], although the gross feature of the calculated dielectric function is quite similar to those of the known ionic compounds.

The linear optical response is directly related to the complex dielectric function $\varepsilon(\omega) = \varepsilon_1(\omega) + i\varepsilon_2(\omega)$. The frequency-dependent dielectric functions for a radiation up to 25 eV are determined using HSE06 ($\alpha$ = 0.4) calculations, as implemented in the VASP package (**Figure S8)** [for details, see ref. [20]], where the symbols *xx*, *yy*, and *zz* refer to the three directions [100], [010] and [001], respectively. The dielectric functions obtained from ABINIT calculations using GGA-ONCV with the scissor operation (fixed at 2.09 eV) are also presented for comparison **(Figure S9)**. The dielectric functions along three main directions show similar values, indicating the nearly isotropic linear optical response of LCPO. The two results are very similar.

The important static dielectric constant $\varepsilon_1(0)$ is given by the zero-frequency limit of $\varepsilon_1(\omega)$. The calculated optical tensor coefficients from different methods are compared in **Table S4**, which reveals that the static dielectric constants $\varepsilon_1(0)$ obtained from HSE06 ($\alpha$ = 0.4) are the smallest among those obtained by different methods. The scissor operation can apparently correct the value of the static dielectric constants. In analogy to the frequency-dependent dielectric functions, the static dielectric matrix $\varepsilon_1(0)$ along different directions show a very small anisotropy. From the calculated frequency-dependent dielectric function from HSE06 ($\alpha = 0.4$), the refractive indices ($n$), birefringence ($\Delta n$) can be obtained, as presented in **Figure S10**. The value of static refractive indices $n^{xx}(0)$, $n^{yy}(0)$ and $n^{zz}(0)$ are 1.563, 1.558, and 1.550, respectively. LCPO exhibits very weak anisotropy, hence low values of birefringence $\Delta n$, which is defined as the maximum differences between the refractive indexes at finite photon energies. The $\Delta n$ is calculated to be 0.013 at the wavelength of 1064 nm (1.17 eV) and it peaks at 16.5 eV in the UV region with a value around 0.25.

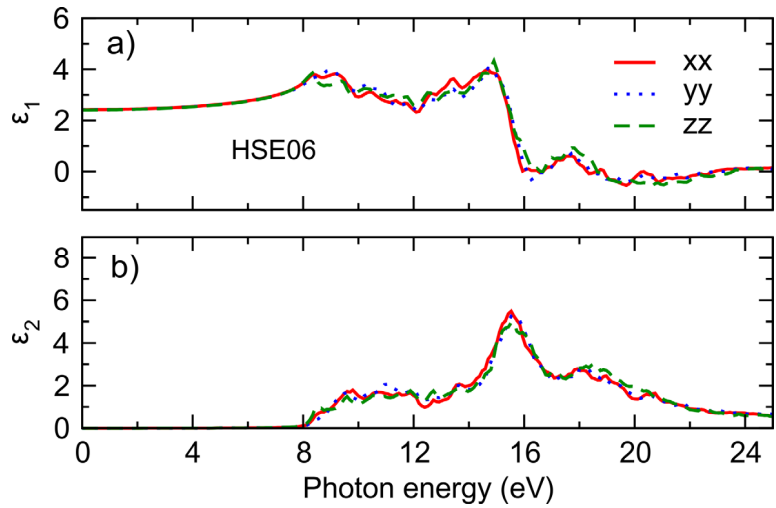

**Figure S8.** Dielectric functions $\varepsilon_1(\omega)$ and $\varepsilon_2(\omega)$ obtained for LCPO by HSE06 ($\alpha = 0.4$) calculations along the three main directions [100], [010] and [001].

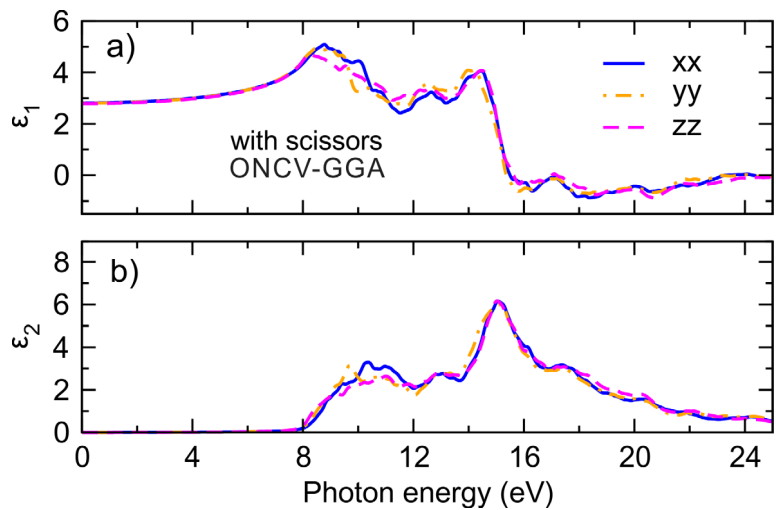

**Figure S9.** Dielectric functions $\varepsilon_1(\omega)$ and $\varepsilon_2(\omega)$ obtained for LCPO by ABINIT calculations (with scissor operation using GGA-ONCV) along the three main directions [100], [010] and [001].

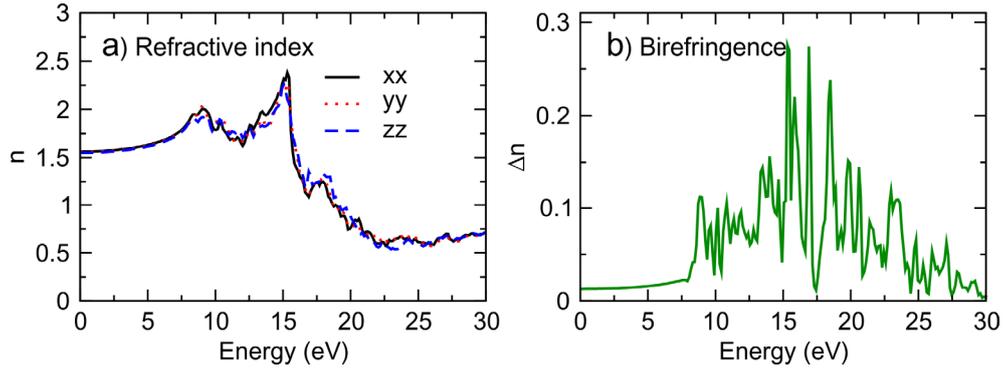

**Figure S10.** Frequency dependent refractive indices $n$ (left) and birefringence $\Delta n$ (right) of LCPO obtained from HSE06 calculations.

**Table S4.** Static dielectric constants of LCPO obtained by using different methods.

| Methods | $\varepsilon_1(xx)$ | $\varepsilon_1(yy)$ | $\varepsilon_1(zz)$ |
|---|---|---|---|
| HSE06 | 2.4440 | 2.4269 | 2.4037 |
| GGA-ONCV (scissor-correction) | 2.8007 | 2.7907 | 2.7752 |
| DFPT (scissor-correction) | 2.7001 | 2.6625 | 2.6414 |
| DFPT (without scissor correction) | 3.0026 | 2.9709 | 2.9356 |
| GGA-PBE (without scissor correction) | 2.9209 | 2.8975 | 2.8798 |

## 6. Second-order polarization of LCPO

### 6.1. Symmetry analysis of the SHG tensor elements $\chi^{(2)}_{opq}$

There are 27 components of the SHG tensor as shown in the following table. Based on the point group of *mm*2, the noncentrosymmetric LCPO (space group No. 36, *Cmc*2$_1$) has seven nonzero SHG coefficients, which are indicated with parentheses.

| $\chi_{111}$ | $\chi_{122}$ | $\chi_{133}$ | $\chi_{123}$ | ($\chi_{113}$) | $\chi_{112}$ | $\chi_{132}$ | ($\chi_{131}$) | $\chi_{121}$ |
|---|---|---|---|---|---|---|---|---|
| $\chi_{211}$ | $\chi_{222}$ | $\chi_{233}$ | ($\chi_{223}$) | $\chi_{213}$ | $\chi_{212}$ | ($\chi_{232}$) | $\chi_{231}$ | $\chi_{221}$ |
| ($\chi_{311}$) | ($\chi_{322}$) | ($\chi_{333}$) | $\chi_{323}$ | $\chi_{313}$ | $\chi_{312}$ | $\chi_{332}$ | $\chi_{331}$ | $\chi_{321}$ |

By crystal symmetry, only the components given with the parentheses can be nonzero.

By the intrinsic symmetry,

$\chi_{113} = \chi_{131}$

$\chi_{223} = \chi_{232}$.

By Kleinman symmetry

$\chi_{311} = \chi_{113} = \chi_{131}$

$\chi_{322} = \chi_{223} = \chi_{232}$

In the Voigt notation, we have

$\chi_{113} = \chi_{131} = 2d_{15}$

$\chi_{311} = 2d_{31}$

$\chi_{223} = \chi_{232} = 2d_{24}$

$\chi_{322} = 2d_{32}$

$\chi_{333} = 2d_{33}$.

Finally, the matrix form of the SHG tensor in the Voigt notation is given by

$$d_{oj} = \begin{pmatrix} 0 & 0 & 0 & 0 & d_{31} & 0 \\ 0 & 0 & 0 & d_{32} & 0 & 0 \\ d_{31} & d_{32} & d_{33} & 0 & 0 & 0 \end{pmatrix} \qquad (2)$$

## 6.2. Sum over states (SOS) method

In this Section we discuss the SHG susceptibility, which is a third-rank tensor $\chi^{(2)}_{opq}(-2\omega,\omega,\omega)$. Note that we use the indices $o, p$ and $q$ instead of the indices $i, j$ and $k$ used in the text for clarity in supplementary text **S6.2** and **S6.3**, which deal with many subscripts and superscripts.

Our work employs the SOS formalism [21] derived by Aversa and Sipe [22] and later modified by Rashkeev *et al.* [23]. In this theory, the SHG susceptibility $\chi^{(2)}_{opq}(-2\omega,\omega,\omega)$ are divided into the interband contribution $_{inter}\chi^{(2)}_{opq}(-2\omega,\omega,\omega)$ and the intraband contribution $_{intra}\chi^{(2)}_{opq}(-2\omega,\omega,\omega)$,

$$\chi^{(2)}_{opq}(-2\omega,\omega,\omega) = {}_{inter}\chi^{(2)}_{opq}(-2\omega,\omega,\omega) + {}_{intra}\chi^{(2)}_{opq}(-2\omega,\omega,\omega) \qquad (3)$$

The interband term is expressed as,

$$_{inter}\chi^{(2)}_{opq}(-2\omega,\omega,\omega) = \frac{e^3}{\hbar^2\Omega}\sum_{vcn,\mathbf{k}}\frac{r^o_{vc}\{r^p_{cn}r^q_{nv}\}}{(\omega_{nv}-\omega_{cn})}\left[\frac{2f_{vc}}{\omega_{cv}-2\omega}+\frac{f_{cn}}{\omega_{cn}-\omega}+\frac{f_{nv}}{\omega_{nv}-\omega}\right] \qquad (4)$$

where the symbols $v$, $c$ and $n$ represent the band indices corresponding to the valence, conduction and intermediate bands, respectively. The intermediate band $n$ can be any of the VBs and CBs, $v \neq c$, and $\mathbf{k}$ is the wave vector. $\{r^p_{cn}r^q_{nv}\} = 1/2\,(r^p_{cn}r^q_{nv} + r^q_{cn}r^p_{nv})$ is a symmetrized combination of the respective dipole matrix elements defined as $r^o_{vc} = p^o_{vc}/im_e\omega_{vc}$, where $p^o_{vc}$ represents a component of the momentum matrix element, $m_e$ is the electron mass. In Eq. 4, $\Omega$ is the normalization volume. The intraband term is written as,

$$_{intra}\chi^{(2)}_{opq}(-2\omega,\omega,\omega)$$

$$= \frac{i}{2}\frac{e^3}{\hbar^2\Omega}\sum_{vc,\mathbf{k}}f_{vc}\left[\frac{2}{\omega_{cv}(\omega_{cv}-2\omega)}r^o_{vc}(r^p_{vc;q}+r^q_{cv;p}) + \frac{1}{\omega_{cv}(\omega_{cv}-\omega)}(r^o_{vc;q}r^p_{cv}+r^o_{vc;p}r^q_{cv})\right.$$

$$\left.+\frac{1}{\omega^2_{cv}}\left(\frac{1}{\omega_{cv}-\omega}-\frac{4}{\omega_{cv}-2\omega}\right)r^o_{vc}(r^p_{cv}\Delta^q_{cv}+r^q_{cv}\Delta^p_{cv}) - \frac{1}{2\omega_{cv}(\omega_{cv}-\omega)}(r^p_{vc;o}r^q_{cv}+r^q_{vc;o}r^p_{cv})\right] \quad (5)$$

with the generalized **k**-space derivative defined as

$$(r^p_{vc})_{;k^o} = \frac{r^o_{vc}\Delta^p_{cv}+r^p_{vc}\Delta^o_{cv}}{\omega_{vc}} + \frac{i}{\omega_{vc}}\sum_n(\omega_{nc}r^o_{vn}r^p_{nc}-\omega_{vn}r^p_{vn}r^o_{nc}) \qquad (6)$$

where $\Delta^o_{vc} = (p^o_{vv}-p^o_{cc})/m_e$ is the difference between the two band velocities. The Fermi factors are 1 if the band is occupied and 0 if the band is empty,

$$f_{vc} = f_v - f_c = 1 \tag{7}$$

$$f_{cn} = H[\varepsilon_n(k) - \varepsilon_F] - 1 = \begin{cases} 0, if\ \varepsilon_n > \varepsilon_F. \\ -1, if\ \varepsilon_n \leq \varepsilon_F. \end{cases} \tag{8}$$

$$f_{nv} = H[\varepsilon_F - \varepsilon_n(k)] - 1 = \begin{cases} -1, if\ \varepsilon_n > \varepsilon_F. \\ 0, if\ \varepsilon_n \leq \varepsilon_F. \end{cases} \tag{9}$$

### 6.3. Effective SHG response $d_{eff}$

To compare with the experimentally measured powder SHG response, the effective $d_{eff}^{2\omega}$ is estimated from the formula derived by Kurtz-Perry [24] and Cyvin *et al.* [25] based on the calculated SHG tensors $d_{abc}^{2\omega}$, i.e., $\frac{1}{2}\chi_{abc}^{2\omega}$,

$$\langle (d_{eff}^{2\omega})^2 \rangle = \frac{19}{105}\sum_a (d_{aaa}^{2\omega})^2 + \frac{13}{105}\sum_{a\neq b} d_{aaa}^{2\omega} d_{abb}^{2\omega} + \frac{44}{105}\sum_{a\neq b}(d_{aab}^{2\omega})^2 + \frac{13}{105}\times \sum_{abc,cyclic} d_{aab}^{2\omega} d_{bcc}^{2\omega}$$

$$+ \frac{5}{7}(d_{abc}^{2\omega})^2 \tag{10}$$

For the point group *mm*2, Eq. 10 can be simplified as,

$$\langle (d_{eff}^{2\omega})^2 \rangle = \frac{19}{105}(d_{333}^{2\omega})^2 + \frac{13}{105}(d_{333}^{2\omega} d_{311}^{2\omega} + d_{333}^{2\omega} d_{322}^{2\omega}) + \frac{44}{105}((d_{113}^{2\omega})^2 + (d_{223}^{2\omega})^2)$$

$$+ \frac{26}{105}(d_{113}^{2\omega} d_{322}^{2\omega}) \tag{11}$$

The effective SHG $d_{eff}$ obtained from different methods for LCPO are listed in **Table 1** in the text. Note that $d_{eff}^{KDP}$ is estimated to be 0.33 pm/V taking $d_{36}^{KDP}$ as 0.39 pm/V.

### 6.4. Comparison of the results from Methods 1 - 5

We now discuss the $d_{eff}$ values obtained from Methods 1 – 5 (**Table 1**). The $d_{eff}$ values of Methods 1 and 5 show the influence of using different pseudopotentials (0.87 vs. 0.48 pm/V), as observed for other compounds [13b, 26]. The results of Methods 4 and 5 show the frequency-dependence of the $d_{ij}$ (0.68 pm/V for ω = 1.17 eV vs. 0.48 pm/V for ω = 0). Method 3 determines the correct bandgap self-consistently, while Methods 1, 2, 4 and 5 do by applying the scissor operation. However, the $d_{eff}$ value of Method 3 does not differ much from those of Methods 1, 2, 4 and 5 (i.e., 0.72 vs. 0.87 – 0.48 pm/V). In calculating the $d_{ij}$ coefficients, Method 2 employs the DFPT while Methods 1, 3, 4 and 5 use the "sum over states (SOS)" method. The $d_{\text{eff}}$ value of Method 2 is not much different from those of Methods 1, 3, 4 and 5 (i.e., 0.79 vs. 0.87 – 0.48 pm/V)..

## 7. Partial response functionals

### 7.1. Definition

We consider a general response function $F(x)$, in which the variable $x$ covers the range of $-\infty < x < +\infty$, and other variables of $F(x)$ are suppressed for simplicity. Then, the partial response functional covering the contribution from the region of $x_0 \leq x < +\infty$ can be defined as

$$\zeta(x_0, F) = \int_{-\infty}^{+\infty} H(x - x_0) F(x) dx \tag{12}$$

where $H(x - x_0)$ is the Heaviside step function (i.e., 0 for $x < x_0$, and 1 for $x \geq x_0$). Note that $x_0$ is the variable of the functional. The partial response functional covering the contribution from the region of $-\infty < x \leq x_0$ can be defined as

$$\zeta(x_0, F) = \int_{-\infty}^{+\infty} \{1 - H(x - x_0)\} F(x) dx \tag{13}$$

The derivative functional $\delta\zeta(x_0, F)$ with respect to $x_0$ provides the signed value of the response function at $x_0$, namely, $\mp F(x_0)$, where the minus and positive sign correspond the cases of Eq.(12) and (13), respectively.

$$\delta\zeta(x_0, F) \equiv \frac{d\zeta(x_0, F)}{dx_0} = \mp \int_{-\infty}^{+\infty} \delta(x - x_0) F(x) dx = \mp F(x_0) \tag{14}$$

When the variable $x$ is a continuous number such as the energy $E$, with the $E_B$ of **Figure 2** acting as the $x_0$, the domain of the variable $E$ can be regarded as $-\infty < E < +\infty$, because $F(E) = 0$ below $E_{min}$ and above $E_{max}$. The variable $x$ can be an integer such as the band index $I$, which runs from 1 for the lowest-lying energy band and increases continuously with increasing the band energy, reaching $N_{tot}$ (i.e., the total number of bands in a given system) for the highest-lying energy band. In this case, the domain of $I$ is $1 \leq I \leq N_{tot}$, and a specific band $I_s$ within the VBs or that within the CBs plays the role of $x_0$.

In the following two sections, each SHG coefficient $\chi_{opq}^{(2)}$ is taken to be the response function $F(E)$ or $F(I)$. We determine the partial response functional $\zeta(E_B, \chi_{opq}^{(2)})$ arising from a partial region of the VBs by using Eq. 12 [denoted as $\zeta_V(E_B)$ in **Figure 2b**], that arising from a partial region of the CBs by using Eq. 13 [denoted as $\zeta_C(E_B)$ in **Figure 2c**]. It should be noted that $\zeta_V(E_B)$ [$\zeta_C(E_B)$] becomes the total response when $E_B$ reaches $E_{min}$ ($E_{max}$).

As discussed above, the partial response functionals can also be discussed in terms of the band index using the specific band $I_s$, leading to the corresponding functionals $\zeta_V(I_s)$ and $\zeta_C(I_s)$ for the VBs and CBs, respectively. There are four derivative partial response functionals $\delta\zeta_V(I_s)$, $\delta\zeta_C(I_s)$, $\delta\zeta_V(E_B)$, and $\delta\zeta_C(E_B)$, which are associated with the partial response functionals $\zeta_V(I_s)$, $\zeta_C(I_s)$, $\zeta_V(E_B)$, and $\zeta_C(E_B)$, respectively. $\delta\zeta_V(I_s)$ provides the contribution of the valence band $I_s$. Similarly, $\delta\zeta_C(I_s)$ provides the contribution of the conduction band $I_s$. Thus, it is possible to calculate the contribution of a specific atom $\tau$ makes to the SHG coefficient from the band state $I_s$ by using its tight-binding atomic orbital representation provided by the VASP code [4b, 4c].

**7.2. Contribution of a specific band to SHG response**

To determine the contribution of a specific band with band index $I_s$ to the SHG coefficients $\chi^{(2)}_{opq}$, we introduce the partial response functionals as a function of $I_s$. Since a given band has information about atomic orbitals, the partial response functionals defined as a function of $I_s$ allow us to determine the contribution of each individual atom to the overall SHG response. To analyze the contribution of a specific band to $\chi^{(2)}_{opq}$, we introduce two partial response functionals as a function of the band index $I_s$, one for the valence bands (VBs), and the other for the conduction bands (CBs). The index $I_s$ increases from 1 to $N_{tot}$ as the band energy $E$ increases from $E_{min}$ to $E_{max}$.

For the VBs, the partial response functional for $\zeta_{opq}(I_s)$ has the intra- and inter-band components. Note that the ω-dependence of $\zeta$ was suppressed for simplicity.

$${}^{VB}_{inter}\zeta_{opq}(I_s) = \sum_v H(v - I_s) \left\{ \frac{e^3}{\hbar^2 \Omega} \sum_{cn,\mathbf{k}} \frac{r^o_{vc}\{r^p_{cn}r^q_{nv}\}}{(\omega_{nv} - \omega_{cn})} \left[ \frac{2f_{vc}}{\omega_{cv} - 2\omega} + \frac{f_{cn}}{\omega_{cn} - \omega} + \frac{f_{nv}}{\omega_{nv} - \omega} \right] \right\} \quad (15)$$

$$\begin{aligned}{}^{VB}_{intra}\zeta_{opq}(I_s) = \sum_v H(v - I_s) &\left\{ \frac{i}{2} \frac{e^3}{\hbar^2 \Omega} \sum_{c,\mathbf{k}} f_{vc} \left[ \frac{2}{\omega_{cv}(\omega_{cv} - 2\omega)} r^o_{vc}(r^p_{vc;q} + r^q_{cv;p}) + \frac{1}{\omega_{cv}(\omega_{cv} - \omega)} (r^o_{vc;q} r^p_{cv} \right.\right.\\
&\left.\left. + r^o_{vc;p} r^q_{cv}) + \frac{1}{\omega^2_{cv}} \left( \frac{1}{\omega_{cv} - \omega} - \frac{4}{\omega_{cv} - 2\omega} \right) r^o_{vc}(r^p_{cv}\Delta^q_{cv} + r^q_{cv}\Delta^p_{cv}) \right.\right.\\
&\left.\left. - \frac{1}{2\omega_{cv}(\omega_{cv} - \omega)} (r^p_{vc;o} r^q_{cv} + r^q_{vc;o} r^p_{cv}) \right] \right\} \end{aligned} \quad (16)$$

For the CBs, the partial response functional for $\zeta_{opq}(I_s)$ also has the intraband and interband components.

$${}^{CB}_{inter}\zeta_{opq}(I_s) = \sum_c [1 - H(c - I_s)] \left\{ \frac{e^3}{\hbar^2 \Omega} \sum_{vn,\mathbf{k}} \frac{r^o_{vc}\{r^p_{cn}r^q_{nv}\}}{(\omega_{nv} - \omega_{cn})} \left[ \frac{2f_{vc}}{\omega_{cv} - 2\omega} + \frac{f_{cn}}{\omega_{cn} - \omega} + \frac{f_{nv}}{\omega_{nv} - \omega} \right] \right\} \quad (17)$$

$$\begin{aligned}{}^{CB}_{intra}\zeta_{opq}(I_s) = \sum_c [1 - H(c - I_s)] &\left\{ \frac{i}{2} \frac{e^3}{\hbar^2 \Omega} \sum_{v,\mathbf{k}} f_{vc} \left[ \frac{2}{\omega_{cv}(\omega_{cv} - 2\omega)} r^o_{vc}(r^p_{vc;q} + r^q_{cv;p}) + \frac{1}{\omega_{cv}(\omega_{cv} - \omega)} (r^o_{vc;q} r^p_{cv} \right.\right.\\
&\left.\left. + r^o_{vc;p} r^q_{cv}) + \frac{1}{\omega^2_{cv}} \left( \frac{1}{\omega_{cv} - \omega} - \frac{4}{\omega_{cv} - 2\omega} \right) r^o_{vc}(r^p_{cv}\Delta^q_{cv} + r^q_{cv}\Delta^p_{cv}) \right.\right.\\
&\left.\left. - \frac{1}{2\omega_{cv}(\omega_{cv} - \omega)} (r^p_{vc;o} r^q_{cv} + r^q_{vc;o} r^p_{cv}) \right] \right\} \end{aligned} \quad (18)$$

As in the case of Eq. 3,

$$^{VB}\zeta_{opq}(I_s) = {}^{VB}_{inter}\zeta_{opq}(I_s) + {}^{VB}_{intra}\zeta_{opq}(I_s) \tag{19}$$

$$^{CB}\zeta_{opq}(I_s) = {}^{CB}_{inter}\zeta_{opq}(I_s) + {}^{CB}_{intra}\zeta_{opq}(I_s) \tag{20}$$

In Eqs. 15-18, $H(v - I_s)$ is a Heaviside step function [27], which projects out the bands with band index lower than $I_s$, while $\{1 - H(c - I_s)\}$ projects out the bands with index larger than $I_s$. Therefore, $^{CB}\zeta_{opq}(I_s)$ is the partial response function for the SHG tensor contributed by the excitations from all VB levels to a specific CB level with band index $I_s$. Similarly, $^{VB}\zeta_{opq}(I_s)$ is the partial response functional for the SHG tensor contributed by the excitations from a specific VB level with band index $I_s$ to all CB levels.

The band index is an integer, so the two functionals, $^{VB}\zeta_{opq}(I_s)$ and $^{CB}\zeta_{opq}(I_s)$, in Eqs. 19-20 are defined in terms of integer numbers. However, one can replace them with real numbers by extending their domain of definition. This extension can be realized, for example, by introducing an interpolation between every neighboring integer pairs. With this analytical extension, $^{VB}\zeta_{opq}(I_s)$ and $^{CB}\zeta_{opq}(I_s)$ can be differentiated classically with respect to the integer argument. Their derivative functionals $\delta\zeta_{opq}(I_s)$ thus obtained indicate uniquely the contribution of a specific band $I_s$ to the SHG tensor component. Therefore, the band-index-dependent functionals for the SHG coefficients $d_{oj}(I_s) = 1/2\,\zeta_{opq}(I_s)$ and the derivative functionals $\delta d_{oj}(I_s)$ can be obtained (See **Figure S11** and **S12**). **Figure S12** shows results by using three different interpolation methods.

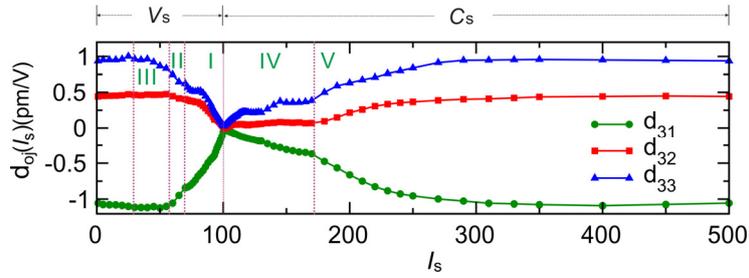

**Figure S11.** Values of the partial response functionals for the SHG components $d_{31}$, $d_{32}$ and $d_{33}$ as a function of $I_s$.

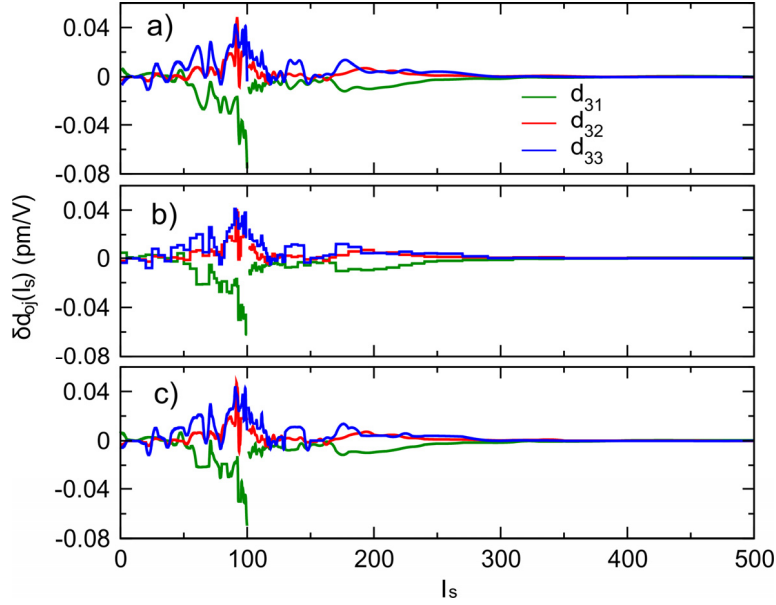

**Figure S12.** Comparison of the derivative functionals $\delta d_{oj}(I_s)$ based on three different interpolation methods, i.e., a) the cubic spline, b) the linear spline, and c) the Akima spline [28].

### 7.3. Individual atom contributions to the SHG response

In the previous section we considered the partial response functionals, $^{VB}\zeta_{opq}(I_s)$ and $^{CB}\zeta_{opq}(I_s)$. A given energy band state with band index $I_s$ is expressed as a linear combination of the atomic orbitals, so one can calculate the contribution of each atomic orbital of each band state. Then, the total contribution of a specific atom $\tau$ can be obtained by summing up the contributions of all the atomic orbitals belonging to the atom $\tau$ from all bands. Since we need to use more indices, to specify these sums, we suppress the indices *opq* and replace $I_s$ with subscript *j* as follows: $^{VB}\zeta_{opq}(I_s)$ to $^{VB}\zeta_j$, and $^{CB}\zeta_{opq}(I_s)$ to $^{CB}\zeta_j$. Namely, $\zeta(I_s)$ is replaced by $\zeta_j$. Suppose that a specific atom $\tau$ has $L$ atomic orbitals with coefficient $^{VB}C_{L\tau}^{\vec{k}j}$ in the valence band $j$ at a wave vector $\vec{k}$. The VASP calculations are carried out in terms of plane wave bases, but the VASP code provides orthonormal atomic orbital bases with which to analyze the computational results. Since these atomic orbitals are orthonormal [4b, 4c], the total contribution $^{VB}A_\tau$ of an atom $\tau$ makes to the SHG coefficient from all the VB bands $j$ is written as

$$^{VB}A_\tau = \frac{\Omega}{(2\pi)^3} \int d\vec{k} \cdot \sum_{L,j} |^{VB}\zeta_j| \left|^{VB}C_{L\tau}^{\vec{k}j}\right|^2 \tag{21}$$

where $\Omega$ is the unit cell volume, the symbol "| |" means the absolute value. Similarly, the total contribution $^{VB}A_\tau$ of an atom $\tau$ makes to the SHG coefficient from all the CB bands $j$ is written as

$$^{CB}A_\tau = \frac{\Omega}{(2\pi)^3} \int d\vec{k} \cdot \sum_{L,j} |^{CB}\zeta_j| \left|^{CB}C_{L\tau}^{\vec{k}j}\right|^2 \tag{22}$$

in which we assumed that the atom has $L$ atomic orbitals with coefficient $^{CB}C_{L\tau}^{\vec{k}j}$ in the conduction band $j$ at a wave vector $\vec{k}$. The $^{VB}A_\tau$ and $^{CB}A_\tau$ values obtained for $\tau$ = Li, Cs, P and O of LCPO are summarized in **Tables S5** and **S6**, respectively. As for the VB contributions to the SHG components $d_{31}$, $d_{32}$ and $d_{33}$, **Table S5** shows that the relative contributions of the individual atoms decrease in the order O >> Cs > P > Li. As for the CB contributions to the SHG components $d_{31}$, $d_{32}$ and $d_{33}$, **Table S6** shows that the relative contributions of the individual atoms decrease in the order Cs > Li > P > O. These results reflect that the SHG of LCPO is governed largely by the occupied orbital states O-2p and Cs-5p as well as by the unoccupied orbital states Cs-5d and Li-2p.

The total contribution, $A_\tau$, each individual atom makes to the SHG response from both the VBs and the CBs (i.e., from all the bands) is given by

$$A_\tau = \frac{\left(^{VB}A_\tau + {}^{CB}A_\tau\right)}{2} \tag{23}$$

where the factor of 1/2 is applied to remove the double counting each excitation. The results of these calculations are summarized in **Table 2** of the text and are discussed there.

This averaging process given in Eqs. 21 and 22 is reasonable when the bands are not very wide. To confirm this point, we also calculated the individual atom contributions to the SHG response by using only one k-point for two different cases. As can be seen from **Figure S4**, the band levels at $\Gamma$ = (0, 0, 0) are quite different from those at in one at Z = (0, 0, 0.5). The individual atom contributions obtained by using only the $\Gamma$ point are very similar to those obtained by using only the Z point. In addition, these two results are very similar to those obtained by using all the k-points of the k-point set; the differences in the relative atom contributions appear only in the second decimal places. Thus, the k-point averaging in Eqs. 21 and 22 is reasonable.

**Table S5.** The contributions of the Li, Cs, P and O atoms to the SHG components $d_{31}$, $d_{32}$ and $d_{33}$ from the valence bands. See **S7.2** for discussion.

| Atom | $d_{31}$ (%) | $d_{32}$ (%) | $d_{33}$ (%) |
|---|---|---|---|
| Li | 1.96 | 2.20 | 1.89 |
| Cs1 | 3.73 | 3.68 | 6.68 |
| Cs2 | 4.18 | 3.88 | 7.28 |
| P | 3.50 | 2.60 | 3.75 |
| O1[a] | 30.40 | 30.27 | 26.35 |
| O2 | 14.29 | 15.54 | 14.04 |
| O3 | 15.65 | 15.24 | 12.97 |

[a] There are two O1 atoms per formula unit.

**Table S6.** The contributions of the Li, Cs, P and O atoms to the SHG components $d_{31}$, $d_{32}$ and $d_{33}$ from the conduction bands. See **S7.2** for discussion.

| Atom | $d_{31}$ (%) | $d_{32}$ (%) | $d_{33}$ (%) |
|---|---|---|---|
| Li | 4.31 | 4.80 | 3.70 |
| Cs1 | 6.81 | 6.67 | 7.74 |
| Cs2 | 6.59 | 6.52 | 7.28 |
| P | 2.51 | 2.47 | 2.21 |
| O1[a] | 3.06 | 3.10 | 3.15 |
| O2 | 1.51 | 1.53 | 1.50 |
| O3 | 1.50 | 1.51 | 1.44 |

[a] There are two O1 atoms per formula unit.

### 7.4. Partial response functionals as a function of energy

We now consider the partial response functionals in terms of energy $\varepsilon_s$ as follows.

$$_{inter}^{VB}\zeta_{opq}(\varepsilon_s) = \sum_v \sum_{cn,\mathbf{k}} H(\varepsilon_{\mathbf{k}v} - \varepsilon_s) \left\{ \frac{e^3}{\hbar^2\Omega} \frac{r_{vc}^o\{r_{cn}^p r_{nv}^q\}}{(\omega_{nv} - \omega_{cn})} \left[ \frac{2f_{vc}}{\omega_{cv} - 2\omega} + \frac{f_{cn}}{\omega_{cn} - \omega} + \frac{f_{nv}}{\omega_{nv} - \omega} \right] \right\} \quad (24)$$

$$_{intra}^{VB}\zeta_{opq}(\varepsilon_s) = \sum_v \sum_{c,\mathbf{k}} H(\varepsilon_{\mathbf{k}v} - \varepsilon_s) \left\{ \frac{i}{2} \frac{e^3}{\hbar^2\Omega} f_{vc} \left[ \frac{2}{\omega_{cv}(\omega_{cv} - 2\omega)} r_{vc}^o(r_{vc;q}^p + r_{cv;p}^q) + \frac{1}{\omega_{cv}(\omega_{cv} - \omega)} (r_{vc;q}^o r_{cv}^p \right. \right.$$

$$+ r_{vc;p}^o r_{cv}^q) + \frac{1}{\omega_{cv}^2}\left(\frac{1}{\omega_{cv} - \omega} - \frac{4}{\omega_{cv} - 2\omega}\right) r_{vc}^o (r_{cv}^p \Delta_{cv}^q + r_{cv}^q \Delta_{cv}^p)$$

$$\left. \left. - \frac{1}{2\omega_{cv}(\omega_{cv} - \omega)} (r_{vc;o}^p r_{cv}^q + r_{vc;o}^q r_{cv}^p) \right] \right\} \quad (25)$$

$$_{inter}^{CB}\zeta_{opq}(\varepsilon_s) = \sum_c \sum_{vn,\mathbf{k}} [1 - H(\varepsilon_{\mathbf{k}c} - \varepsilon_s)] \left\{ \frac{e^3}{\hbar^2\Omega} \frac{r_{vc}^o\{r_{cn}^p r_{nv}^q\}}{(\omega_{nv} - \omega_{cn})} \left[ \frac{2f_{vc}}{\omega_{cv} - 2\omega} + \frac{f_{cn}}{\omega_{cn} - \omega} + \frac{f_{nv}}{\omega_{nv} - \omega} \right] \right\} \quad (26)$$

$$_{intra}^{CB}\zeta_{opq}(\varepsilon_s) = \sum_c \sum_{v,\mathbf{k}} [1 - H(\varepsilon_{\mathbf{k}c} - \varepsilon_s)] \left\{ \frac{i}{2} \frac{e^3}{\hbar^2 \Omega} f_{vc} \left[ \frac{2}{\omega_{cv}(\omega_{cv} - 2\omega)} r_{vc}^o (r_{vc;q}^p + r_{cv;p}^q) + \frac{1}{\omega_{cv}(\omega_{cv} - \omega)} (r_{vc;q}^o r_{cv}^p \right.\right.$$

$$\left. + r_{vc;p}^o r_{cv}^q) + \frac{1}{\omega_{cv}^2} \left( \frac{1}{\omega_{cv} - \omega} - \frac{4}{\omega_{cv} - 2\omega} \right) r_{vc}^o (r_{cv}^p \Delta_{cv}^q + r_{cv}^q \Delta_{cv}^p)\right.$$

$$\left.\left. - \frac{1}{2\omega_{cv}(\omega_{cv} - \omega)} (r_{vc;o}^p r_{cv}^q + r_{vc;o}^q r_{cv}^p) \right] \right\} \tag{27}$$

Based on Eqs. 24-27, the SHG coefficients $d_{oj}(\varepsilon_s) = 1/2\, \zeta_{opq}(\varepsilon_s)$ and their derivatives with respect to energy $\delta d_{oj}(\varepsilon_s)$ can be obtained.